\documentclass[preprint]{aastex}
\usepackage{epsfig}
\usepackage{enumerate}
\def\beq{\begin{equation}}
\def\eeqno#1{\label{#1}\end{equation}}

\def\rarrow{\rightarrow }

\def\dleft{\rlap{{\it D}}\raise 8pt
\hbox{$\scriptscriptstyle\Leftarrow$}}
\def\dright{\rlap{{\it
D}}\raise 8pt\hbox{$\scriptscriptstyle\Rightarrow$}}

\def\kms{{\rm km~s^{-1}}}

\def\az{a_{0}}

\def\l0{\ell_{0}}

\def\rar{\rightarrow}
\def\s{\sigma}
\def\a{\alpha}
\def\b{\beta}
\def\l{\lambda}

\def\c{\gamma}
\def\f{\phi}

\def\r{\rho}
\def\m{\mu}
\def\n{\nu}
\def\z{\zeta}

\def\vinf{V\_{\infty}}

\def\CC{\textbf{C}}

\def\A{\mathcal{A}}

\def\L{\mathcal{L}}
\def\O{\mathcal{O}}

\def\D{\Delta}
\def\Df{\D \f}
\def\d{\delta}

\def\drt{d^3\vr}
\def\a{\alpha}
\def\xlimin{{x\rarrow\infty \atop{\raise 1pt\hbox to 30pt
{\rightarrowfill}}}}
\def\limlim#1#2{{#1\rarrow #2 \atop{\raise 1pt\hbox to 30pt
{\rightarrowfill}}}}

\def\vr{{\bf r}}
\def\vn{{\bf n}}

\def\vd{{\bf d}}

\def\vF{{\bf F}}
\def\vv{{\bf v}}

\def\vg{{\bf g}}
\def\vvt{{\bf v}(t)}
\def\vrt{{\bf r}(t)}

\def\va{{\bf a}}

\def\vf{{\bf f}}
\def\vF{{\bf F}}

\def\S{\Sigma}

\def\grad{\vec\nabla}
\def\div{\vec \nabla\cdot}
\def\gf{\grad\phi}

\def\Q{\mathcal{Q}}
\def\L{\mathcal{L}}
\def\fN{\f^N}
\def\gN{g\_N}

\def\gmn{g_{\m\n}}

\def\m{\mu}
\def\a{\alpha}
\def\b{\beta}
\def\c{\gamma}
\def\d{\delta}

\def\n{\nu}

\def\D{\Delta}
\def\fpg{4\pi G}

\def\_#1{_{\scriptscriptstyle #1}}
\def\^#1{^{\scriptscriptstyle #1}}

\def\sm{\S\_M}
\def\azg{\A_0}
\def\gz{g_0}
\def\MM{M\_M}
\def\vgz{\textbf{g}_0}
\def\RM{r\_M}
\def\vsav{\langle V^2\rangle}
\def\rp{\r\_p}
\begin{document}
\title{MOND laws of galactic dynamics}
\author{Mordehai Milgrom}
\affil{Department of Particle Physics and Astrophysics, Weizmann Institute}

\begin{abstract}
MOND predicts a number of laws that galactic systems should obey irrespective of their complicated, haphazard, and mostly unknowable histories -- as Kepler's laws are obeyed by planetary systems. The main purpose of this work is to show how, and to what extent, these MOND laws follow from only the paradigm's basic tenets: departure from standard dynamics at accelerations  $a\lesssim\az$, and space-time scale invariance in the limit $a\ll \az$. Such predictions will be shared by all MOND theories that embody these premises.
This is important because we do not know which of the existing MOND theories, {\it if any}, is a step in the right direction.
In the Newtonian-dynamics-plus-dark-matter paradigm, the validity of such clear-cut laws -- which tightly constrain baryons, `dark matter', and their mutual relations -- is contrary to expectations.

\end{abstract}
\maketitle
\section{Introduction}
MOND is an alternative paradigm of dynamics, replacing Newtonian dynamics (ND) plus dark matter (DM) in the nonrelativistic (NR) regime, and general relativity (GR) in the relativistic regime (Milgrom 1983, for an extensive, recent review of MOND see Famaey \& McGaugh 2012). The basic tenets of NR MOND are: (1) A new constant, $\az$, with the dimensions of acceleration, is introduced into dynamics. (2) A MOND theory is required to reduce to standard dynamics in the limit of accelerations\footnote{Here, and in many other instances, when I refer to `accelerations' in comparison with $\az$, I mean `all quantities with the dimensions of acceleration'. This is analogous to relativity reducing to ND when all quantities with the dimensions of velocity are much smaller than the speed of light (not only velocities, but also, e.g. the square root of the gravitational potential).} much larger than $\az$ -- formally effected in a given theory by taking the  limit $\az\rar 0$. (3) In the opposite, deep-MOND limit (DML), of accelerations $g\ll\az$ everywhere -- formally effected by taking the limit $\az\rar \infty, ~~G\rar 0$, with $\azg\equiv G\az$ fixed (see details in section \ref{DML}) -- the theory has to become space-time scale invariant (SI), namely invariant under $(t,\vr)\rar\l(t,\vr)$.
\par
The original formulation of the DML in Milgrom (1983) posited a relation between the MOND acceleration, $g$, and the Newtonian acceleration, $g\_N\sim MG/r^2$, of the form $g\sim(\az g\_N)^{1/2}$, which satisfies tenet (3) (both sides scaling as $\l^{-1}$). However, the third tenet as formulated above is precise and general, and should thus be used instead of the rough relation between accelerations, which cannot hold exactly, in general.
\par
We can liken the role of $\az$ in MOND to that of $\hbar$ in quantum mechanics, or to that of $c$ in relativity: all these constants, first, delineate the validity domain of the old, classical theory, in which they do not appear. Secondly, they appear saliently in disparate phenomena in the revised domain.\footnote{As we shall see below, it is not $\az$ itself that appears in deep-MOND phenomena: it is always $\azg$ that does.} MOND tenet (2) is then analogous to the correspondence principle in quantum dynamics, for $\hbar\rar 0$, and to relativity tending to ND for $c\rar\infty$.
\par
Beyond the basic tenets, one wishes to construct an NR theory of dynamics based on them, and then extend the theory to a replacement of GR. There are several examples of such theories. In the NR regime, we have the suitably chosen nonlinear extension of the Poisson equation (Bekenstein \& Milgrom 1984), and a quasilinear MOND formulation (QUMOND; Milgrom 2010a). These are classified as `modified gravity' (MG) theories (more on these in Sec. \ref{dila}).\footnote{Generalizations of these can be constructed. For example, for two-potential theories, a Lagrangian can be written as \cite{milgrom10a}: $\L=\frac{1}{2}\r \vv^2(\vr)-\r\f({\vr})-\L_f[(\gf)^2,(\grad\psi)^2,\gf\cdot\grad\psi)]$
that embody the MOND tenets. They have a DML of the form
\beq \L_f=\azg^{-1}\sum_{a,b} s\_{ab}[(\gf)^2]^{a+3/2}[(\grad\psi)^2]^{a+b(2-p)/2}(\gf\cdot\grad\psi)^{b(p-1)-2a}, \eeqno{iiaa}
where $p$ is fixed for a given theory, the third tenet is satisfied for any set of $a,~b$.
The dimensions of $\f$ and $\psi$ are, respectively, $[l]^{2}[t]^{-2}$ and (for $b\not = 0$) $[l]^{2-p}[t]^{2(p-1)}$; thus, by our convention, discussed below in Sec. \ref{DML}, their scaling dimensions are $0$ and $p$. The coefficients $s\_{ab}$ are dimensionless.
 For any $p$,
the nonlinear Poisson theory is gotten with $a=b=0$, while QUMOND is the special case with $p=-1$ and two terms with $a=-3/2,~b=1$  and $a=-b=-3/2$.}
There are also `modified inertia' (MI) formulations, for which there is not yet a full fledged theory, but, on whose implications much can be said nonetheless (Milgrom 1994, 2011). For more details on these and on relativistic MOND theories see the review by Famaey \& McGaugh (2012).
\par
The departures of quantum theory and of relativity from the classical pictures that preceded them were not mere changes in form of the equations of classical dynamics. They each brought totally new concepts to underlie dynamics. This may well turn out to be the case for an eventual deeper MOND theory. A hint of such eventuality may be seen in the possible connections that MOND brings out between local dynamics and the global state of the Universe (see Milgrom 2009a and references therein).
\par
Existing MOND theories are extensions of the pre-MOND dynamics (NR and relativistic) that introduce $\az$, add new degrees of freedom (DoFs), and modify the underlying action; but, they do not deeply depart in spirit from their predecessors.\footnote{Such theories might be analogous to the pre GR attempts to relativitize  gravity by writing Lorentz invariant theories for the gravitational potential -- as, e.g., in Nordstr\"{o}m's theories -- before the full force of gravity-as-space-time-geometry has been appreciated.} Perhaps one of them will turn out to be an effective theory that captures the essence of the deeper MOND theory.\footnote{The appearance in all  these theories of an interpolating function is an indication that, at best, they can be effective theories.} But perhaps none of the existing MOND theories will turn out to be a step in the right direction; perhaps their merits are summed up in their being  self-consistent theories that embody the basic tenets of MOND.
\par
It is thus noteworthy that one can predict a number of general laws, based almost entirely on the basic premises. These laws are important to recognize and understand in detail for several reasons: (a) They constitute the core predictions of MOND, obeyed by all MOND theories, present and future, that embody the basic premises.
(b) They focus attention on well-defined, easy-to-grasp, and memorable sub-predictions of MOND, subsumed by the general prediction that MOND would make of the general dynamics of a system. (c) They involve only limited information about the systems (such as global attributes) and hence are easier to test on large samples [in contrast with, e.g., rotation-curve (RC) analysis, which requires knowledge of the full RC, and the detailed baryon mass distribution]. (d) They constitute a `to-derive' list for the competing paradigm of ND-plus-DM. (e) Pinpointing and quantifying the remaining differences in these laws as predicted by different theories will help discern between theories.
\par
Succinctly formulated, most of the laws recognized to date are as follows.
\begin{enumerate}[(i)]
\item{\label{asym}}
Speeds along an orbit around any bounded mass, $M$, become asymptotically independent of the size of the orbit (being dependent only its shape). For example, for circular orbits   $V(r\rar\infty)\rar\vinf(M)$.
This replaces Kepler's third law $T^2=K(M)r^3$ with $T=\bar K(M) r$.
\item{\label{btfr}}
$\vinf(M)=(MG\az)^{1/4}$, which dictates the dependence of $\bar K$ on the central mass.
\item{\label{bound}}
A mass discrepancy appears when we cross $a=\az$. In disc galaxies, the transition from `baryon dominance' to `DM dominance' occurs always around the radius where  $V^2(r)/r=\az$.
\item{\label{siglimit}}
Quasi-isothermal systems, which well model many galactic systems, have mean surface densities
$\bar\Sigma\lesssim \sm\equiv\az/
2\pi G$.
\item{\label{msigma}}
Quasi-isothermal, or deep-MOND, systems of mass $M$, have characteristic velocity dispersion $\sigma\sim (MG\az)^{1/4}$.
\item{\label{efe}}
The external field in which a system is falling affects its intrinsic dynamics: the external-field effect (EFE).
\item{\label{disc}}
Disc galaxies behave as if they have both disc and
spherical `DM' components.
\item{\label{stability}}
MOND endows self gravitating systems with an increased, but limited, stability.
\item{\label{max}}
The incremental acceleration MOND predicts (that attributed to `DM') can never much exceed $\az$.
\item{\label{centralsig}}
The central surface density of `dark haloes' is $\lesssim \sm$.
\end{enumerate}
\par
It is reassuring that even the more elaborate prediction of RCs in MOND follows almost entirely from the basic premisses alone.
\par
The above-listed laws have been discussed before (Milgrom 2008,2009a, Famaey \& McGaugh 2012). But, in the following sections I consider in great detail their meaning, implications, and how they are derived. And, in particular, I will show to what extent, and how, these laws follow from the basic tenets alone, and what aspects of them are theory dependent or require additional assumptions.
\par
Some generalities to take home concerning these laws are as follows.
\newline
a. To a large extent they do follow from only the basic tenets, and are thus shared in one way or another, by all MOND theories that embody these tenets.
\newline
b. Several of them  revolve around $\az$ in different roles.
\newline
c. These laws are independent as phenomenological laws.\footnote{In the same sense that without the unifying framework of quantum dynamics, the different quantum phenomena -- such as the blackbody spectrum, the photoelectric effect, the hydrogen-atom spectrum, superconductivity, etc. -- would appear unrelated phenomena that somehow involve the same constant $\hbar$.} For example, if interpreted as effects of DM in ND, one can construct model families of baryons plus DM that will satisfy any subset of these laws but not the others.
\newline
d. Some of these laws, when interpreted in terms of DM, would describe properties of the `DM' alone [e.g., (\ref{asym}), (\ref{max}), and (\ref{centralsig})], of the baryons alone [e.g., (\ref{siglimit})], or relations between the two [e.g., (\ref{btfr}), (\ref{bound}),  and (\ref{disc})].
\par
Comparisons of these predicted laws with data are discussed at length in Famaey \& McGaugh (2012) and in many previous studies, and will not be considered here.
\section{\label{additional}Additional assumptions}
\par
In deriving the MOND laws, I assume that MOND does not involve additional dimensioned constants.\footnote{Counter examples are the MOND theories discussed in Babichev \& al. (2011), which adds a length scale, and in Zhao \& Famaey (2012), based on a suggestion in Bekenstein (2011), which adds a velocity constant.} I also assume that it does not involve dimensionless constants that much differ from unity, $k,...$, whose appearance is tantamount to having in the theory several, greatly different, acceleration constants, $\az,~k\az,~k^{-1}\az,$ etc., that may play different phenomenological roles. My whole discussion below hinges on the premise that this does not happen.\footnote{TeVeS (Bekenstein 2004), and its NR limit, is a counterexample: it does have additional dimensionless parameter(s). TeVeS does not tend to GR in the mere limit $\az\rar 0$, and to make it agree with various high-acceleration constrains, one of these constants is forced to be much different from unity.}
This assumption means, in particular, that the DML (with its SI) applies already below a relatively large fraction of $\az$. It also implies that ND holds approximately already above a few $\az$. I will call this latter `weak compatibility' of MOND with standard dynamics.\footnote{The second tenet constitutes a compatibility requirement, but it requires compatibility only in the very limit $\az\rar 0$.} In some special cases one also assumes `strong compatibility': the requirement that MOND approaches standard dynamics very fast.\footnote{This can be defined in various quantitative way, the exact form of which is immaterial here. For example, existing MOND theories all involve some interpolating function between the Newtonian and MOND regimes, of the form $\m(A/\az)$, where $A$ is a quantity with the dimensions of acceleration. Weak compatibility means, for these theories, that $\m(x)\sim 1$ already beyond $x\sim$ a few. Strong compatibility may be taken as the requirement that $[1-\m(x)]\ll p/x$ for $x\gg 1$ ($p$ is some constant of order unity).} Such strong  compatibility is hardly ever relevant to galactic dynamics, which only probe accelerations up to $g\sim10\az$. However, dynamics in systems such as the Solar-system and binary pulsars, do probe accurately much higher accelerations.
\par
More assumptions are usually made implicitly, and I spell them out here. It is assumed, for example, that, as in ND, all the (NR) effects of a body at distances much larger than its size depend only on the mass of the body, not, e.g., on its internal structure, or on attributes other than its total mass.\footnote{Namely, finite size effects are assumed to decay with distance faster than the dominant contribution that depends only on the total mass.}  This means, for example, that predicting the dynamics of a system made of bodies small compared with their separations requires only knowledge of their masses. I shall refer to such bodies in such a context as pointlike. This assumption can be shown to hold in the known NR MOND theories, such as the nonlinear Poisson theory (Bekenstein \& Milgrom 1984), QUMOND (Milgrom 2010a), and the NR limits of all the relativistic MOND theories discussed to date. But, it might be that a future MOND theory will violate this assumption (e.g., by adding coupling constants other than the mass). This requires , in particular, that the bodies and systems we are dealing with have non-zero masses. There are theories and situations, where the effective gravitating mass of a system might vanish, as happens, e.g., in the NR limit of bimetric MOND (BIMOND), for a system of equal amounts of matter and twin matter (see Milgrom 2010b for a detailed discussion). In this case, which requires special treatment, and which I bar in this paper, the asymptotic field does depend on the structure of the system even if it is very small in extent (see below).

\section{\label{DML}The deep-MOND limit}
The DML is pivotal in understanding the MOND laws; so I expand here on its properties.
The DML may be formally described as follows: Apply a space-time scaling to all the DoFs in the equations of motion of a theory, corresponding to $(t,\vr)\rar\l(t,\vr)$, and let $\l\rar\infty$.\footnote{The DoFs may have nonzero scaling dimensions; for example, a scalar field may transform as $\psi(\vr,t)\rar\l^\a\psi(\vr/\l,t/\l)$, where $\a$ is the scaling dimension of $\psi$. Note the appearance of $\l$ in the denominator for the independent variables -- a possible source of confusion.}
If the limit exists as a consistent theory, it is automatically SI (because scaling again by a finite factor $\l$ has no further effect).  In such a limit, all accelerations scale as $g\rar\l^{-1}g\rar 0$, and so this also corresponds to a situation where $g\ll\az$ (note that the constants of the theory, such as $G,~\az$ are not affected by the scaling). If the limit is also nontrivial -- in the sense that it retains the physics we want to describe -- the theory is a good candidate MOND theory. For example, for a theory with only $G$, possibly $c$, and masses as constants, the limit can only describe a theory with zero masses.\footnote{For example, ND has such a limit that is SI, but the Poisson equation becomes $\Df=0$.} It is the presence of $\az$ that allows us to get such an interesting SI limit for purely gravitational theories.\footnote{In a theory that includes electromagnetism, charges (and currents) $e\rightarrow \lambda^{-1/2}e\rightarrow 0$, such that $e^2 a_0$ remain fixed in the limit, and electromagnetic interactions remain finite in a SI theory.}
\par
An equivalent definition of the limit is to leave the DoFs alone, but scale the constants according to their dimensions, such that if a constant $q$ has dimensions $[q]=[l]^a[t]^b[m]^c$, then $q\rar \l^{-(a+b)}q$; then let $\l\rar\infty$. This equivalence is seen by noting that all equations are unchanged by a change of units; so if we apply a scaling by a factor $\l$ to the DoFs, and then change the units (which affects the values of the DoFs and the constants), so as to leave the values of the DoFs intact,\footnote{The DoFs can be normalized, e.g., by multiplying them by a power of $\az$, so that their scaling dimension matches their $[l][t][m]$ dimensions: if $[\psi]=[l]^\b[t]^\c[m]^\z$, then the scaling dimension is $\a=\b+\c$. I assume this normalization everywhere.} the constants suffer the above scaling. In our context, in a pure gravity theory, involving only $c,~G,~\az$, and masses $m_i$, we have $c\rar c$, $G\rar \l^{-1}G$, $\az\rar\l\az$, and $m_i\rar m_i$. So, $\az\rar\infty$, as befits the DML, $G\rar 0$, but $c$, $m_i$, and $\azg\equiv G\az$ remain fixed, and they are the only constants that can remain in a DML.\footnote{ For example, the DML of the nonlinear Poisson formulation of MOND can be written as
$\va=-\gf$, with $\div[|\gf|\gf]=4\pi\azg\r$, where $\f$ has dimensions of $[l]^2[t]^{-2}$, and indeed it has $\a=0$; this is the desired normalization. But we could also work with $\psi\equiv\f/\az$, with the equations taking the equivalent form $\va=-\az\grad\psi$, with $\div[|\grad\psi|\grad\psi]=(4\pi G/\az)\r$. Now, $\psi$ has dimensions $[l]$ but must still have $\a=0$ to retain SI, not consistent with its dimensions. Indeed we see that with this normalization, $\az$ and $G$ do appear separately, not as $G\az$. In MI theories, we have schematically, $a\va/\az=-\gf$, with $\Delta\f=\fpg\r$. To have SI, the first equation implies that $\f$ has to have $\a=-1$, which is incongruent with its dimensions -- and indeed $G$ and $\az$ appear separately. Working with $\psi\equiv\az\f$, for which the scaling dimension is the desired one, we have  $a\va=-\grad\psi$ with $\Delta\psi=4\pi \azg\r$.}
\par
In a universe governed strictly by a DML theory, there is neither $\az$ nor $G$, only $\azg$ appears. If it were not for the existence of a Newtonian regime of observed phenomena, we would not have known of $G$, and would not have spoken of $\az$. Below, I shall use $\azg$ in deep-MOND results.\footnote{In ND, $G^{1/2}$ has to be introduced as a dimensioned proportionality factor between the inertial ($m$) and the gravitational mass ($m\_G$). The former is defined so that inertial forces are $ma$, and the latter is defined such that gravitational forces scale as $m\_G^2/r^2$; so $m\_G$ has the dimensions of $G^{1/2}m$. (The masses, $m$, we use all along are the inertial ones.) In the DML, gravitational forces must scale as $p(m,\az,G)/r$ to retain SI. Dimensionally, we must have $p\sim \azg^{1/2}m^{3/2}$; so gravitational forces are then $F\sim \azg^{1/2}m^{3/2}/r$.} It is also useful sometimes to
write DML results in terms of the MOND mass $\MM\equiv c^4/\azg$ (Milgrom 2008,2009a)

\par
DML dynamics was originally (Milgrom 1983) epitomized by the relation $g\sim(\az\gN)^{1/2}$ between the MOND and Newtonian accelerations of test particles. However, except for special circumstances,\footnote{For example, for 1D configurations in some theories.} the MOND acceleration cannot simply be a function of the local Newtonian acceleration. Such an algebraic relation does not hold exactly in any known MOND theory, and is also not consistent with conservation laws. In MI theories the MOND acceleration, unlike $\vg\_N$, is not even a function of position, but depends also on the full trajectory.  So, unlike the formulation of the DML here, by the requirement of SI, which is exact and clear-cut in meaning, the $g-\gN$ relation is neither. Still, this algebraic relation is very useful, and does apply under some circumstances, where it can be derived from the above basic assumptions, as we shall see.

\par
Now that we know that only $\azg$ appears in the dynamics, we infer (e.g., Milgrom 2009a) that the theory is invariant, more generally, to all scalings that do not change the value of $\azg$ (when considered as change of units): This is a two-parameter family  $l\rar\a l$, $t\rar\b^{-1} t$, $m\rar (\a\b)^4 m$. This specific two-parameter invariance comes about because NR MOND involves only the constant $\az$ in addition to $G$, together with its SI.\footnote{Had we built MOND, for example, around a new length constant, $\ell$, instead of an acceleration, and require SI in the limit of large distances, $\ell\rar 0$, then only the ratio $G/\ell$ would have appeared in the limit. In this case the two-parameter family of invariances would have had $m\rar (\a\b)^2 m$.}
Thus, if $\vr_i(t)$ is a system history for $m_i$, then   $\a\vr_i(\b t)$, with velocities $\a\b\vv_i(\b t)$, is a system history for masses $(\a\b)^4 m_i$ (with appropriately scaled initial conditions).\footnote{For continuum systems, if the density and velocity fields $\r(\vr,t),~\vv(\vr,t)$ are a solution, so are $\a\b^4\r(\vr/\a,\b t),~\a\b\vv(\vr/\a,\b t)$.}
\par
Imagine then an isolated DML system, in virial equilibrium, of total mass $M$ made of test particles of masses $m_i=Mq_i$, and of characteristic size
$R$, and characteristic internal speed $V$; so we can write for the particle speeds $\vv_i=V\omega_i$, with some average of the
$\omega_i$ (say their mass weighted rms) being 1.
Define
\beq \Q=\frac{V^2}{(M\azg)^{1/2}}=\frac{V^2/R}{(\az MG/R^2)^{1/2}}=\frac{g}{(\az \gN)^{1/2}}, \eeqno{nasfara}
where $g\equiv V^2/R$ is what we would identify with the characteristic MOND acceleration, and $\gN\equiv MG/R^2$ is the characteristic Newtonian acceleration. It follows from the above that every such system is a member of a two-parameter family of systems having all values of $M$ and $R$ (provided they are still in the DML; i.e., with $MG/R^2\ll\az$), but all having the same value of $\Q$. In other words, the value of $\Q$ for a DML system cannot depend on any of the dimensioned attributes of the system ($M,~R,~V$), only on dimensionless and SI attributes,\footnote{So, $\Q$ cannot depend on the dimensionless, but non-SI $MG/R^2\az$.} such as $q_i$, $\omega_i$, shape parameters, orbit distribution, etc. all of which are the same within each family of systems. For example, DML, pure-exponential discs of all masses and sizes are in the same family; all DML isothermal spheres (ISs) with the same constant anisotropy ratio are in the same family, etc.
\par
While $\Q$ is predicted to be strictly constant within each family, the above arguments alone do not prevent it from varying greatly among the families. Indeed, many of the dimensionless parameters on which $\Q$ might depend, in principle, can take up values much smaller or much larger than unity, and very different among the families. Still, I argue that {\it due to our extra assumptions}, $\Q$ cannot vary greatly among sub-$\az$ systems, and is, in fact, of order unity for all of these.
\par
The argument hinges strongly on my above-stated assumption that MOND does not involve dimensionless constants very different from unity.\footnote{As opposed to dimensionless system parameters, which can, of course, take up small or large values.} In particular, it follows that the transition from the DML to the almost Newtonian regime is relatively narrow, and always occurs within a factor of a few $\az$ in acceleration.
Take then a system with $MG/R^2=\az$. The above assumption says that its dynamics is not much different from Newtonian, so, with our definition of $V$, it satisfies approximately the Newtonian virial relation $MG/R\sim V^2$. Eliminating $R$ gives
$\Q\sim 1$ for such systems. But, for each DML family of scaled systems -- which fills the region (much) below the borderline parabola $M=(\az/G)R^2$ in the $R-M$ plane -- there are members near the borderline. By the above assumption, the dynamics of these members is not very different from that of their neighbours on the borderline that have a similar geometry, mass distribution, and other dimensionless attributes.\footnote{For example, all DML, pure-exponential discs have a similar $\Q$ value to borderline exponential discs; all DML ISs with the same constant anisotropy ratio have their borderline counterparts, etc.} So they too satisfy $\Q\sim 1$, and so do all the DML systems.\footnote{It is easy to see that if, for example, contrary to our assumption, the DML (i.e., the region of SI) applies only up to accelerations $k\az$ ($k\ll 1$), above which the theory transits quickly to Newtonian behaviour, then DML systems have $\Q\sim k$; and, in principle, $k$ could even depend on system type.}
We have thus derived the approximate DML relation $g\sim (\az\gN)^{1/2}$ between the characteristic {\it global} parameters of a system from the basic tenets and our additional assumptions. Such a relation does not necessarily apply locally.
\par
The meaning and the expected dynamics of the relativistic DML are not understood yet: because it turns out that $\az\approx cH_0/2\pi$ (Milgrom 1983); ($H_0$ is the Hubble constant), a system that is both relativistic, and of accelerations $a\ll\az$ has to have a size larger than the Hubble distance; so this double limit does not apply to any known system.
\subsection{\label{dila}Space-dilatation invariance in modified-gravity theories}
Additional, important DML predictions follow from the basic tenets in the framework of the large class of MG theories.
\par
A relativistic, MG MOND theory is a metric theory where the matter action, $S\_M$, is the standard one, while the Einstein-Hilbert action for the metric is replaced by a modified action, $S\_G(\gmn,A_s,c,G,\az)$, which may involve additional DoFs, $A_s$.
For purely gravitational systems made of masses $m_p$ -- as we consider here -- the matter action is
$-\sum_p m_pc^2\int d\tau_p$. In the NR (NR) limit of such a theory the matter action becomes $\int \sum_p m_p\{\frac{1}{2}\vv^2_p(t)-\f[\vr_p(t)]\}~dt$.
We can, as usual, separate the problems of solving the motion of the masses, governed by the unmodified $\ddot\vr_p=-\gf$, and that of solving, at any time, for the gravitational fields, given the mass distribution at this time: $\r(\vr,t)=\sum_p m_p\d^3[\vr-\vr_p(t)]$, treated as a given external source. The latter problem, which we now concentrate on, is governed by the NR action  $S=\int L_g dt$, where,
 \beq L_g=-\int\r(\vr)\f(\vr)\drt-
L_f(\f,\psi_a,G,\az), \eeqno{kalur}
with $t$, now only a parameter, not a variable, suppressed.
Here, I use the continuum description of the mass distribution, $\f$ is the NR gravitational potential -- the only gravitational DoF that couples directly to matter -- and $\psi\_a$ are the other gravitational DoFs. The gravitational-field Lagrangian, $L_f$, is a functional of $\f$ and $\psi\_a$; it involves $G$ and $\az$ as the only dimensioned constants.
Consider now the DML of such a problem. The special form of all MG theories, which all retain the matter equation of motion $\dot\vv=-\gf$, requires for SI that the scaling dimension of $\f$ is $0$. With our standardized normalization of the other DoFs, $L_f\rar L\^{DML}_f(\f,\psi_a,\azg)$.
\par
Scale invariance of the DML implies that $L_g$, and separately $L_f$, is invariant to space dilatations $\vr\rar\l\vr$, if we keep the scaling dimensions of $\f$ and the $\psi$s as their dilatation dimensions (I reserve `scaling' for space time, and `dilatation' for space alone). This is because the equations of motion derived from $L_g$ do not involve time derivatives and hold for each time separately. We can renormalize all DoFs by powers of $\azg$ so that they have no time dimensions (i.e. $[\psi]=[m]^\a[l]^\b$); so, for example, use $\hat\f=\azg^{-1/2}\f$. Since then the only quantity on which $L_f$ depends that has time dimensions is $\azg$, and since $L_f$ itself has dimensions $[m][l]^2[t]^{-2}$, we must have $L_f=\azg^{1/2}\hat L_f(\f,\psi_f)$. Thus, $\azg$ disappears from the problem altogether, since
$L_g=-\azg^{1/2}(\int\r\hat\f+\hat L_f)$.
The only dimensioned constants that remain are masses.
\par
This dilatation invariance of $L_g$, which follows from only the basic tenets for NR limits of MG theories is, in itself, a powerful result. For example, it was shown in Milgrom (2013b) that it leads to a general virial relation
 \beq\sum_p \vr_p\cdot\vF_p=-\frac{2}{3}\azg^{1/2}[(\sum_p m_p)^{3/2}-\sum_p m_p^{3/2}],   \eeqno{bery}
which holds for an isolated DML system of pointlike masses, $m_p$, at positions $\vr_p$, and subject to gravitational forces $\vF_p$.
From this follow various important applications, such as an analytic expression for the DML two-body force, and a general velocity-mass relation (see Sec. \ref{massigma}). Relation (\ref{bery}) had been derived in the special cases of the nonlinear Poisson (Milgrom 1997) and the QUMOND (Milgrom 2010a) theories. But it is now known to hold for any MG theory, and to follow from only the basic tenets applied to such theories.
\subsection{Nonlinearity of the deep-MOND limit}
All existing MOND theories have a nonlinear NR DML.\footnote{Relativistic versions are nonlinear also in the high-acceleration regime, as GR is.}
Is this forced by the basic tenets? The global relation
$g\sim (\az\gN)^{1/2}$, which we derived from our assumptions, and, e.g., law (\ref{btfr}), which follows from the basic tenets, imply already that we cannot superpose accelerations produced by several mass distributions, as is possible in ND (DML accelerations scale as the square root of the mass). But this does not necessarily imply that the theory is nonlinear. A linear NR dynamics is one where the motion of a test particle induced by a mass distribution $\r_1+\r_2$ can be gotten by superposing the motions calculated for $\r_1$ and $\r_2$ separately.
This means, more precisely, that the particle equations of motion of all particles are of the form
 \beq (\O\vr_i)(t)=(\CC\r)(\vr_i), \eeqno{wusda}
where $\O$ is a linear (differential, possibly time-nonlocal) operator acting on the trajectory $\vr_i$ to give another vector function of time, ($\O\vr_i)(t)$, and $\CC$ is a vector, linear operator acting on the field $\r$. Only the constants $G$ and $\az$ may appear in the two operators. However, since $\r$ appears linearly and only on the right hand side, $G$ can only appear as its prefactor. Also, for a DML theory we can only have $\azg$ appearing. This means that $\CC=\azg\bar\CC$, where $\bar\CC$, like $\O$, cannot contain any dimensioned constants. In particular, since we do not have constants with dimension of time, $\O$ must be $\propto d^n/dt^n$.
(Time cannot appear explicitly because of time-translation invariance.) Assume, for simplicity, that $\bar\CC$ does not involve time derivatives. Since $\azg\r$ has dimensions $[l][t]^{-4}$, $\bar\CC$ can have no length dimensions, and $n=4$. (If we do allow time dimension $-p$ for $\bar\CC$, with a $p$th time derivative appearing therein, we must have $n=4+p$.) We then must have from the basic tenets
 \beq \frac{d^4\vr_i(t)}{dt^4}=\azg\bar\CC\r(\vr), \eeqno{wusdala}
where $\bar\CC$ is dimensionless. This is schematically tantamount to replacing the standard, nonlinear DML relation $a^2\sim \azg M/r^2$ by the linear $\ddot a\sim\azg M/r^3$, which retains SI.

As one example of many, we could take
 \beq \bar\CC\r=\int\frac{\r(\vr')
 (\vr-\vr')d^3\vr'}{|\vr-\vr'|^4}=\grad\psi(\vr),  \eeqno{pasaf}
with
\beq \psi(\vr)=-\frac{1}{2}\int\frac{\r(\vr')
 d^3\vr'}{|\vr-\vr'|^2}.  \eeqno{pasafal}

(The field $\psi$ is not a modified gravitational potential, since $\grad\psi$ is not the acceleration.)
\par
Note that in a theory of this type both the kinetic and gravity terms are modified so it is neither a MG theory, nor a pure MI theory.\footnote{The equation of motion (\ref{wusdala}), with the choice (\ref{pasaf}), can be gotten from a Lagrangian $\L\propto \sum_i [(d^2\vr_i/dt^2)^2-2\azg\psi(\vr_i)]$.}
\par
Equation (\ref{wusdala}) is the most general equation of motion describing a linear DML theory (for $p=0$). It is, however, not an acceptable DML theory:
a. It is a high order theory, with all the known associated problems: the awkwardness of having to dictate four initial conditions, the inevitable presence of instabilities as implied by Ostrogradski's theorem (Ostrogradski 1850), discussed in detail in Woodard (2007), and the related problem of exploding solutions.
b.  There is no general MOND theory that has this equation of motion as its DML, and reduces to ND in the limit $\az\rar 0$. Clearly there is no kinetic action that interpolates between $(d^2\vr_i/dt^2)^2$ and the Newtonian $(d\vr_i/dt)^2$ and involve only $\az$. It was shown generally in  Milgrom (1994) that there is no local kinetic Lagrangian (one depending only on time derivatives of $\vr_i$), with Galilei invariance, that has both a correct DML and a Newtonian limit.
\par
Thus, if we exclude theories of this type, nonlinearity is forced  by the basic tenets. Still, such a DML formulation may be useful as a toy test bed in various contexts.

\section{MOND laws of galaxy dynamics}
I now discuss the MOND laws in detail: how they are derived, and to what degree they follow from the basic tenets.
Some of these laws can be described in terms of the properties of a putative DM distribution that reproduces MOND effects. However, not all MOND theories have an equivalent description in terms of DM.
In MG theories, which only modify the gravitational potential from the Newtonian $\fN$ to the MOND potential $\f$, the modification is fully described by the (Newtonian) gravitational effects of `phantom matter' (PM)\footnote{This is a very useful concept in the context of MOND. For some applications see Milgrom (1986), Milgrom \& Sanders (2008), Wu \& al. (2008), and Zhao \& Famaey (2010).} of density
\beq \rp=\frac{1}{\fpg}(\D\f-\D\fN).   \eeqno{lakio}
But, this is not the case, for example, in MI theories where the modification may depend not only on position, but also on the particle orbits.
\par
Still, when analysing the dynamics of galactic systems, the data have always been limited, never relying on studies of all orbit types, at all positions. With such limited information on the dynamics of the system, a description in terms of DM equivalent may be possible even when it does not apply to the full dynamics. I refer here to the properties of `DM' obtained in this limited way.  For example, RC analysis pertains only to the dynamics of circular orbits in the equatorial plane of an axisymmetric disc galaxy. With this limited data, the departure from ND can be attributed to a spherical halo of DM, even in MOND theories where this would not be possible for general orbits.

\subsection{\label{asympt}Asymptotic constancy of the orbital velocity around an isolated mass}
In MOND, the orbital speeds become independent of the size of the orbit for an orbit far outside some isolated mass. This follows simply from SI of the DML: under scaling, the size of the orbit and all the relevant times are multiplied by the same factor; so velocities do not change (the scaling of the central mass distribution is immaterial for asymptotically far orbits).
\par
More formally, consider some isolated mass distribution, $\r(\vr)$, of total mass $M$, having a characteristic radius $R$ (some minimal radius containing most of the mass in the system). Take $\vrt$ to be some asymptotic, test-particle orbit. By `asymptotic' in the MOND context one understands two conditions: a. $|\vrt|\gg R$; so the orbit is far outside the mass, which can then be considered a point mass; so, the dynamics of such orbits are oblivious to the exact form of $\r$ or its extent $R$. b. $|\vrt|\gg\RM$, where $\RM\equiv (MG/\az)^{1/2}$ is the MOND radius of the system; so the orbit is wholly in the DML. In particular, the orbital dynamics does not change if we expand the system so that $R\gg\RM$ (but still $|\vrt|\gg R$). This puts the whole system is in the DML, and we can apply SI: For any asymptotic orbit $\vr(t)$, the orbit $\hat\vr(t)=\l\vr(t/\l)$ is an orbit for the distribution $\hat\r=\l^{-3}\r(\vr/\l)$, also having total mass $M$. But, since the structure of $\hat\r$ is immaterial, we conclude that $\hat\vr$ is an orbit for $\r(\vr)$ itself. The velocities on the scaled orbit are $\hat\vv(t)=\vv(t/\l)$. Thus, for such asymptotic orbits the orbital velocities are independent of the size of the orbit.
\par
This is an obvious result for logarithmic potentials, but I have shown that it follows, more generally from the basic tenets.
For circular orbits, this means that for $r\gg R,~\RM$, the rotational speed becomes independent of $r$,\footnote{I assume, that the theory is not so odd as to permit orbits with the same radius, but different velocities.}
a result that is exact for isolated systems in all MOND theories.\footnote{Interesting departures occur when the theory allows gravitating masses of opposite signs, as e.g., in BIMOND with twin matter (Milgrom 2010b). In a system with vanishing total mass, the asymptotic speeds are not constant: Here, the dilatation of $\r(\vr)$ itself does affect the asymptotic field (the asymptotic field is no more the canonical MOND logarithmic potential). For example, in NR BIMOND, the field equation was solved exactly in Milgrom (2010b) for two opposite masses $\pm m$ in the DML. Asymptotically, the potential is
$\f\approx -(m\azg)^{1/2}\vr\cdot\vd/r^2$,
where $\vd$ is the dipole separation. So, asymptotic speeds decrease as $(m\azg)^{1/4}(d/r)^{1/2}$. The breakdown of the general result occurs because the asymptotic potential is not invariant to the dilatation of the mass distribution (under which $\vd\rar \l\vd$).
However, masses of opposite signs repel each other, so one cannot have a self gravitating system of this type. In what follows I shall ignore such systems.}
\par
For comparison of this prediction with data see the many published compilations of RCs, summarized in Fig. 15 of Famaey \& McGaugh (2012, see also their Figs. 21-27), as well as Milgrom (2012a) for two elliptical galaxies, and Milgrom (2013a) for a weak-lensing test of all galaxy types.
\par
If interpreted in terms of DM, this prediction pertains to a pure-halo property, since asymptotics is governed by the strongly dominant putative DM halo.
\subsection{The mass-asymptotic-velocity relation}
Given that the asymptotic dynamics of test particles depend only on the total mass, and given that the only quantity with dimensions of velocity that can be constructed from mass, length, and $\azg$ is $(M\azg)^{1/4}$, the orbital speed on any asymptotic trajectory has to be of the form
 \beq \vvt=(M\azg)^{1/4}\vec \z(t)=c\left(\frac{M}{\MM}\right)^{1/4}\vec \z(t), \eeqno{malamala}
where $\vec \z(t)$ covers all dimensionless orbits, which are independent of the specific MOND theory, of details of the mass distribution, or of the size of the orbit (they are the dimensionless classical orbits in a logarithmic potential). In particular, for circular (constant speed) orbits,
$|\vec \z(t)|$ is a universal constant for the theory. The common
convention is to normalize $\azg$ (or $\az$, given $G$), such that for circular orbits
$|\vec \z(t)|=1$. Thus, the mass-asymptotic-velocity relation
 \beq M=\azg^{-1}\vinf^4=\MM\left(\frac{\vinf}{c}\right)^4,  \eeqno{datr}
between the total central mass and the asymptotic circular velocity, is an exact prediction of the basic tenets of MOND, and of all the theories that embody them.
\par
For recent comparisons with data see, e.g., McGaugh(2011a,2012), and Figs. 3-4 of Famaey \& McGaugh (2012), as well as Milgrom (2012a) for two elliptical galaxies, and Milgrom (2013a) for a weak-lensing test of all galaxy types.
\par
If interpreted in the context of DM, this prediction constitutes a relation between the total baryonic mass and a halo property, the asymptotic speed.

\subsection{The mass discrepancy correlates with acceleration, and in disc galaxies appears always at a fixed acceleration value}
In disc galaxies for which $a\equiv V^2(r)/r>\az$ in the inner parts, the mass discrepancy is predicted by the basic tenets to appears always around the radius where $a=\az$: Because the only quantity with the dimensions of acceleration in this context is $V^2(r)/r$, the transition from Newtonian to MONDian dynamics, dictated by the basic tenets, must occur at a radius where $V^2(r)/r=\az$, if such a radius exists. Systems in which $V^2/r<\az$ everywhere are predicted to show MONDian behaviour everywhere, with no transition.
A similar behaviour is expected in existing theories also for quasi-spherical, pressure-supported systems.
\par
For comparison with data see, e.g., Fig. 12 in Scarpa (2006) for pressure-supported systems, and Fig. 10 of Famaey \& McGaugh (2012) for disc galaxies, both showing clearly that the mass discrepancy is a function of $a/\az$ with (gradual) onset at $a=\az$.
\par
If interpreted within the DM paradigm, this prediction constitutes a tight connection between visible matter and DM, since it predicts where DM dominance takes over from visible-matter dominance in a any system.

\subsection{Quasi-isothermal spheres have mean surface densities
$\langle\S\rangle\lesssim \sm\equiv \az/
2\pi G$}
Self-gravitating ISs of finite mass (possibly with anisotropic velocity distributions) do not exist in ND: if the mass of a self gravitating body is finite, then beyond a certain radius its enclosed mass is nearly constant. Beyond this radius, Newtonian speeds of bound particles have to decrease as $r^{-1/2}$, which is inconsistent with isothermality.\footnote{Viewed differently, for an IS, the Newtonian equation of equilibrium is $\hat\r\equiv dln(\r)/dln(r)\propto -M(r)/r$. Beyond the radius containing most of the mass, $\hat\r$ thus tends to zero, while finiteness of the mass would imply that asymptotically $\hat\r<-3$.}
In MOND, however, asymptotic isothermality for a finite mass is natural, as we saw in Sec. \ref{asympt}. It follows that for a self gravitating sphere to exist and be protected from evaporation (by MOND), it cannot be deep in the high-acceleration regime; namely, it cannot have a size $R\ll \RM$. But $R\gtrsim \RM$ means that the mean surface density $\langle\S\rangle\sim M/\pi R^2\lesssim \sm$.\footnote{Non-ISs, such as stars, can, of course, have arbitrarily high values of $\langle\S\rangle$.}
\par
The above results follow from only the basic tenets of MOND, but $\S_M$ should be taken as only a rough upper limit on, and an accumulation value of, $\langle\S\rangle$.
MOND ISs have been considered in detail for the nonlinear Poisson formulation in Milgrom (1984). (These results apply also in QUMOND, since the two theories coincide for spherical systems.)
It was shown there (Fig. 2) that IS families with different anisotropy ratios have upper limits $\langle\S\rangle<\a \sm$, with $\a\sim 1$ for isotropic orbits, becoming somewhat larger for more tangential orbit, and becoming smaller, even as small as $\a=0.1$ or smaller for more radial orbits.
\par
Inasmuch as globular clusters, dwarf spheroidal galaxies, elliptical
galaxies, and galaxy clusters are quasi-isothermal (not all individual systems within each class are approximately IS), we can apply this prediction to them. Indeed they satisfy this inequality, as is shown, e.g., by the `Fish law' for ellipticals, and see Fig. 6 of Scarpa (2006). This prediction might even be pertinent to giant molecular clouds (Milgrom 1989b).
\par
This law, if interpreted within the DM paradigm pertains to pure baryonic attributes.

\subsection{\label{massigma} A mass-bulk-velocity relation}
For isolated, virialized, self-gravitating systems with $\langle\S\rangle\lesssim\sm$, the basic tenets of MOND predict a correlation:
\beq \s^4\sim M\azg=c^4\left(\frac{M}{\MM}\right), \eeqno{sisista}
where $M$ is the total mass, and $\s$ is the characteristic, bulk velocity.\footnote{Defined, for example, as the mass-weighted RMS velocity $\s^2=\vsav=M^{-1}\int \r(\vr)v^2(\vr)~\drt$.}
This follows from the results of section \ref{DML} identifying $\s$ with $V$ there.
\par
Despite the similar appearance, this law is very different from
law (\ref{btfr}), $\vinf^4=M\azg$: (a) The latter is an exact relation, the former only a correlation with possibly large scatter. (b) The latter involves the asymptotic, circular speed, the former a bulk, characteristic speed within the system. (c) The latter applies to all isolated bodies, the former is predicted for limited classes.\footnote{For example, for an {\it isolated} Sun-like star, which is highly Newtonian in its bulk ($\langle\S\rangle\gg\sm$), law (\ref{btfr}) should still hold, with $\vinf\approx 0.4\kms$. However, the characteristic bulk velocity is $\s\approx (\RM/R)^{1/2}\vinf\sim 400\kms$, so clearly the mass-bulk-velocity relation is not satisfied.}
\par
Quasi-ISs are a relevant special case, since by law (\ref{siglimit}) they have $\langle\S\rangle\lesssim\sm$ -- they are either DML or borderline cases -- and so are all predicted to satisfy correlation (\ref{sisista}). Seen differently, in ISs the bulk $\s$ is, by definition, also the asymptotic $\s$. But the latter must be $\sim \vinf$ for the system. Thus, in this case the $M-\s$ correlation does follow from law (\ref{btfr}).
\par
Clearly, there is considerable range that is allowed for the value of $\Q=\s^2/(M\azg)^{1/2}$ by the arguments leading to correlation (\ref{sisista}). But this range in the ratio is small compared with the relative spread of values of $\s^2$ or $M^{1/2}$ separately, within the gamut of systems to which this prediction applies -- from globular clusters and dwarf spheroidals, through galaxies of all types, to clusters of galaxies.
\par
From relation (\ref{bery}), which holds for any DML system of pointlike masses, $m_i$, in any MG MOND theory, one can derive the value of $Q$ (e.g., Milgrom 1997)
\beq \Q=\frac{2}{3}(1-\sum q_i^{3/2}), \eeqno{varavar}
where $q_i=m_i/M$ are ratios of $m_i$ to the total mass $M$, and $\s$ is the mass-weighted root-mean-squared (3-D) speed.\footnote{Note that once we define the pointlike bodies, this choice enters both the masses, $m_i$, and the definition of $\s$. Internal velocities within the bodies are then not included in the calculation of $\s$. If $m_i$ themselves are made of DML constituents, we can write a $Q$ value for the system of all constituents, but the $\s$ value of this system is different.}
For a smooth mass distribution (made of `test particles', for which $\sum q_i^{3/2}\ll 1$) we get a universal value $\Q=2/3$ for any such DML system irrespective of its structure. This prediction has been tested, for example in small galaxy groups (Milgrom 2002), and, recently, by predicting successfully the internal velocity dispersions of many dwarf satellites of the Andromeda galaxy (McGaugh \& Milgrom 2013a,b).
 \par
We do not have such a general result for MI theories. However, for such theories there is a simple prescription for calculating the RC of any axisymmetric disc galaxy (Milgrom 1994). Taking advantage of this, I calculated (Milgrom 2012b) the $\Q$ value for large classes of thin DML disc galaxies, where only the rotational speeds contribute to $\s$ (defined as above). I found, remarkably, that all have $\Q$  values within a narrow range: $\Q\approx0.73\pm0.01$.
\par
ISs with a constant but arbitrary velocity-anisotropy ratio were considered in Milgrom (1984) for the full MOND (not only DML), nonlinear Poisson theory (the results apply also to the later QUMOND). It was found that for all such spheres $2/3\le\Q<1$, where the exact value depends on the anisotropy ratio, and on the closeness to the DML.
\par
In the toy, linear DML theory described by eq.(\ref{wusdala}) with the choice (\ref{pasaf}) -- which satisfies the third tenet -- one can derive an exact relation of the form
 \beq \langle a^2\rangle=\azg M^{-1}\langle\sum_{i<j}\frac{m_i m_j}{|\vr_i-\vr_j|^2}\rangle_t, \eeqno{klama}
for a steady-state system, where $\langle a^2\rangle$ is the time average, mass-weighted system average of the squared acceleration, and $\langle\rangle_t$ signifies time average. If we replace the sum with $\sim M^2/R^2$, and  $\langle a^2\rangle$ with $\sim \s^4/R^2$, we get $Q\sim 1$ for this `theory'.
\par
This MOND law underlies the observed Faber-Jackson relation for elliptical galaxies, and its generalization to all pressure-supported systems, as discussed by Sanders (2010). He applied it to systems spanning almost 10 orders in M and $\s^4$, and also extended the discussion to a three-parameter fundamental plane, allowing for an additional parameter to $M$ and $\s$. For additional comparison with data see  Famaey \& McGaugh (2012). In addition to the intrinsic scatter expected around relation (\ref{sisista}), there is scatter due to conversion of line-of-sight dispersions to 3-D ones, to other deviations from the definition of $\s$ used in our derivation, to departures from assumptions in the derivation (e.g., isothermality), to errors in the masses (which have to be converted from luminosities), etc.
\par
When interpreted in the DM paradigm this prediction pertains sometimes to pure-baryon properties (when $\langle\S\rangle\sim\sm$, so baryons dominate), and sometimes to a relation between baryons and `DM' (when $\langle\S\rangle\ll\sm$, so `DM' dominates).
\subsection{\label{EFE}The external-field effect}
Unlike ND, MOND is nonlinear even in the NR regime. It generally does not satisfy the strong equivalent principle; so effects of an overall acceleration on the internal dynamics of a system are generically expected. To be able to say what constraints the basic tenets impose on such effects I have to confine myself here to theories whereby only the instantaneous value of the external acceleration matters. This excludes from the discussion a large class of MI theories that are time nonlocal (Milgrom 1994). In these, the full (external) trajectory of the system enters, which complicates the discussion. Some of the possible consequences of such nonlocality are discussed briefly in Milgrom (2011), but what follows here does not apply to such theories.
\par
Consider then a system of mass $m$ (`system $m$'), and extent $r$, that is falling in the field of a mother system with acceleration whose instantaneous value is $\vgz$ . Assume that the theory and conditions are such that, to a good enough approximation, all the information about the mother system enters the dynamics within $m$ only through $\vgz$. One can then write
 \beq \va=\va(m,r,\az,G,g\_0,\vn\_0,\a),  \eeqno{mudas}
where $\va$ stands for the {\it internal} acceleration runs of elements of $m$, namely the full acceleration in the field of the mother system minus $\vgz$ (suppressing the dependence on position, time, and particle index). It is written as a function of all the available dimensioned independent parameters, as well as of $\vn\_0$, the unit vector in the direction of $\vgz$, and of $\a$, which stands for the many dimensionless parameters that characterize the configuration, such as all the mass ratios, and all the geometrical parameters (angles, ratios of all distances to $r$, etc.).
Here I am only interested in scaling laws of the dimensioned parameters -- for example, in how $|\va|$ depends on the dimensioned system attributes -- so I shall suppress the dependence on $\vn\_0$ and $\a$.
\par
Since $\va/g\_0$ is dimensionless, it can depend only on dimensionless quantities; so we can write, most generally
\beq \va=g\_0\vF^*(\eta,\theta),~~~~~~  \eta\equiv \frac{mG}{ r^2\gz}\sim \frac{g\_N}{\gz},~~~~~~\theta\equiv \frac{g\_0}{\az}. \eeqno{mudasi}
When $\gz\ll|\va|$, its effects can be neglected. So here I shall be interested in the opposite case, of external-acceleration dominance, $\gz\gg|\va|$.\footnote{It is difficult to make general statements about the intermediate case where the two accelerations are of the same order.} The above choice of dimensionless variables is useful for this case.
Clearly, $\vF^*(0,\theta)=0$. So we are interested in the behaviour of $\vF^*$ to lowest order in $\eta$. We shall see below that external-field dominance requires $\eta\ll 1$ when $\theta\gtrsim 1$, and the SI condition $\eta\ll\theta$ when $\theta\lesssim 1$ (in which case the whole problem is in the DML; $\eta$ and $\theta$ both scale like $\l^{-1}$ under scaling); so we can write this condition generally as $\eta\ll min(1,\theta)$.
\par
We do not know that a MOND theory is necessarily expandable in powers of $\eta$ near $\eta=0$. But assuming that it does, I write
 \beq \va\approx \gz\eta^q \vf(\theta), ~~~~~~\eta\ll min(1,\theta). \eeqno{mudasii}
 (I assume that $q$ does not depend on $\theta$; see below.)
\par
To constrain $q$ and $\vf(\theta)$ I now employ the basic tenets of MOND. The limit $\az\rar 0$, namely when
$\az\ll|\va|\ll\gz$, is strongly Newtonian for all accelerations, and is within the validity domain of eq.(\ref{mudasii}). For this region, $g\_0$ and $\az$ have to disappear from expression (\ref{mudasii}). This implies that $q=1$, and that $|\vf(\theta\gg 1)|\sim 1$, such that $(mG/r^2)\vf(\infty)$ is the internal Newtonian acceleration field within $m$.
This means that the internal dynamics is Newtonian for any value of $g\_N$ when $\theta\gg 1$; i.e., also when $g\_N\ll\az$. In other words: whenever the external field is highly Newtonian and dominates over the internal field, the latter is necessarily Newtonian. This result holds also when $q$ depends on $\theta$, because then we still must have $q(\theta\rar\infty)\rar 1$.
\par
More generally, inasmuch as $q=1$ for all $\theta$, we can write eq.(\ref{mudasii}) in its full validity domain (external-field dominance) as
 \beq \va=\frac{mG}{r^2}\vf(\theta). \eeqno{dasac}
This means that when the external field is dominant, the internal dynamics is always quasi Newtonian, in the sense that the accelerations scale as $mG/r^2$, only with an enhanced effective constant $G_{eff}\sim G|\vf(\theta)|$, and with not-quite-Newtonian geometrical aspects that stem from the fact that $\vf$ has different geometric properties than $\vf(\infty)$: for example, $\vf$ depends on the direction relative to $\vn\_0$, and on the theory at hand, while $\vf(\infty)$ does not.
\par
When $\theta\ll 1$ the whole system is in the DML, where the basic tenets dictate that eq.(\ref{dasac}) becomes SI. Under scaling, $\theta$ scales like $\gz$, namely $\theta\rar\l^{-1}\theta$ (since $\gz$ is a DML acceleration of the mother system it scales as $\gz\rar\l^{-1}\gz$). This means that $\vf$ must become proportional to $\theta^{-1}$: $\vf(\theta\ll 1)\approx\theta^{-1}\bar\vf$. We see then that $\vf(\theta)$ has the same asymptotic behaviours as $1/\m(\theta)$, where $\m$ is the interpolating function appearing in present MOND theories.
\par
If $q$ does depend on $\theta$, the EFE does not conform to the standard results. For example, in the DML we could have $0<q(0)\not = 1$, in which case SI dictates $\vf(\theta\ll 1)\approx \theta^{-q(0)}\hat\vf$. Then $a\sim \gz(g\_N\az/g\_0^2)^{q(0)}=\gz(\eta/\theta)^{q(0)}$. We see that, as stated above, the condition for external-field dominance, $a\ll\gz$, when $\theta<1$, is indeed always $\eta\ll\theta$.
\par
For example, if $q(0)=1/2$, this gives the standard scaling of the MOND acceleration in isolated systems $a\sim (g\_N\az)^{1/2}$; i.e., there is no EFE, except for effects in $\bar\vf(0)$. So the basic tenets lead to the standard EFE results (indeed to an EFE) only if some additional analytic properties are assumed.
\par
The toy DML theory described by eq.(\ref{wusdala}), which satisfies scale invariance (but which does not combine with an appropriate Newtonian limit), does not lead to an EFE.
\par
The above analytic assumptions do
hold in all the MOND formulations considered to date: e.g., in the original, pristine formulation in (Milgrom 1983), in the formulation of Bekenstein \& Milgrom (1984), and in QUMOND (Milgrom 2010a).
For example, in QUMOND, we can write schematically (ignoring the vectorial nature of the quantities involved)
 \beq a/g\_0\sim\n[\theta\m(\theta)+\theta\eta][\m(\theta)+\eta]-1,  \eeqno{shpsh}
where $\n(y)$ is the QUMOND interpolating function, and $\m(x)$ is such that $\n[x\m(x)]\m(x)=1$. We have $\m(\theta\ll 1)\approx \theta$, $\m(\theta\gg 1)\approx 1$; so we see explicitly why the condition $\eta\ll min(1,\theta)$ is tantamount to a dominant external field. And, clearly the next to zeroth-order term is $a/\gz\sim \eta(1+\hat\n)/\m(\theta)$, where $-1/2<\hat\n<0$ is the logarithmic derivative of $\n$.
\par
In summary, the fact that an external field $|\vgz|\gg\az$ renders the internal dynamics Newtonian, follows from only the basic tenets of MOND, provided that only the instantaneous external field enters the internal dynamics (not necessarily true in MI, time-nonlocal theories). This is relevant, for example, to experimental results in the laboratory and Solar system, and to the dynamics of star clusters near the sun. On the other hand,
the specific form of the EFE when $|\vgz|\ll\az$, even its very existence, is not strictly dictated by the basic tenets alone. Its basic features do follow under another plausible assumption, shared by all full-fledged theories considered to date: that the expansion power in eq.(\ref{mudasii}) does not depend on $\theta$.
\par
There is no EFE in the DM paradigm.
\subsection{Disc galaxies have both disc and a halo components of `phantom matter'}
The disc component of a spiral galaxy is described ideally as a thin planar mass. In ND, the surface density of the disc is related to the perpendicular component of the measured acceleration just outside the disc, by $\S=(2\pi G)^{-1}g\^{out}\_{N\perp}$ (assuming symmetry to reflection in the disc plane). MOND predicts a perpendicular component of the acceleration $g\^{out}\_{M\perp}\not = g\^{out}\_{N\perp}$. This difference would be interpreted by a Newtonist as being due to a thin disc of `PM' of surface density
 \beq \S=(2\pi G)^{-1}(g\^{out}\_{M\perp}-g\^{out}\_{N\perp}), \eeqno{mopyta}
just as the general-field difference between the MOND accelerations is interpreted as `halo of PM'. Such a phantom disc is predicted only where there is a baryonic disc, because MOND does not create a discontinuity where one is not created by the mass distribution.
More generally, it would appear to a Newtonist that a certain distribution of disc DM is needed to explain the perpendicular (z) dynamics within the disc.
\par
This effect has been discussed in specific formulations of MOND starting from the first papers on MOND (see, in particular, Milgrom (2001) -- where observational indications of disc `DM' are also discussed -- and Famaey \& McGaugh (2012) with further references therein; possible peculiarities of the effect in nonlocal, MI theories were alluded to in Milgrom (2011)). It is difficult to make general statements based only on the basic tenets. However, under some circumstances, the effect can be treated similarly to the EFE treated in Sec. \ref{EFE}. Consider the perpendicular dynamics up to some height $Z$ at galactocentric position $r$. If the disc were a uniform planar structure, we could have treated the $z$ dynamics as a 1D problem with planar symmetry. This might be a good approximation at the very centre of the disc, provided we explore it to $Z\ll h$, where $h$ is the typical scale length of the disc. In ND, this is a reasonable approximation at all $r$, provided we limit ourselves to $Z\ll r$. MOND, however does not generally permit such a treatment away from the centre, even for $Z\ll r$: the radial component of the dynamics at $r$ must also be reckoned with because of the nonlinearity of MOND. As in the case of the EFE, if we can assume that the only effects of the $r$ dynamics on the $z$ dynamics enter through the value of $g_r(r)$, the value of the radial acceleration at $r$ -- which can be assumed a constant as regards the $z$ dynamics at $r$ -- then our treatment of the EFE applies here with $|\vgz|$ replaced by $g_r(r)$.
\par
A disc component of DM, as predicted by MOND, is not a natural occurrence in the cold-dark-matter paradigm.

\subsection{MOND endows self-gravitating systems with added, but limited stability}

By ND, accelerations in a self gravitating system scale as $a\sim \r R G$ (where $R$ is size and $\r$ density). In the DML, SI (and dimensional considerations) dictates that
$a\sim (\r R\azg)^{1/2}$. So, while in ND $\d a/a\propto \d \r/\r$, in the DML we have $\d a/a\propto (1/2)\d \r/\r$: the response to perturbation is weaker, by a factor $\sim2$, leading to higher, but not much higher, stability.
\par
In other words, MOND predicts added stability in DML systems compared with systems of similar structure that are in the Newtonian regime. But, it also predicts that the degree of stability does not depend on how deep in the MOND regime the systems is. This follows from the invariance of the DML to the two-parameter family of scalings discussed in Sec. \ref{DML}. Such scalings relate systems with all possible mass discrepancies, namely with all possible values of $MG/R^2\az$ (as long as they are in the DML), which must then all have the same stability properties. For example, all exponential discs, with the same ratio of scale height to scale length, the same ratio of vertical velocity dispersion to rotational speeds, etc., but having all possible values of $M/R^2\ll\az/G$, are predicted by the basic tenets of MOND to have the same stability properties. What differs among them is the time scale (for instabilities to grow, for example).
\par
Detailed discussions of disc stability in specific MOND theories can be found in Milgrom (1989a), Christodoulou (1991), Brada \& Milgrom (1999b), and Tiret \& Combes (2007a,b).
Observational implications are summarized in Famaey \& McGaugh (2012).
\par
The predicted MOND dependence of disc stability properties on the mass discrepancy (as evinced by the RC) is not reproduced in the DM paradigm: There, a progressively large RC mass discrepancy bespeaks a progressively more massive halo compared with the disc. This, in turn, predicts progressively higher disc stability, unless one allows a large fraction of the putative DM to reside in a disc component, which is not natural in the cold-dark-matter paradigm. In fact, the observations of DML galaxies, with large mass discrepancies, and yet showing signs of instability (such as spiral structure and bars) is a challenge for this paradigm.
\subsection{\label{maximum}The excess MOND accelerations
 (those putatively attributed to DM) cannot much exceed $\az$}
In regions of a system in which $\gN\lesssim\az$, the excess MOND accelerations\footnote{The Newtonian acceleration is always well defined; the `MOND acceleration' not always so: In MG MOND theories the notion of an acceleration field is well defined in the NR case as $-\gf$, where $\f$ is the MOND potential dictated by the theory. In MI theories, where the actual acceleration of a test particle may depend on details of its orbit, there is no `acceleration field'. However, as things stand today, at least, one always imposes the existence of such a field when interpreting the data. See more details in Sec. \ref{maxac}.} cannot much exceed $\az$. If they would, the total acceleration, $g$, would be much above $\az$, the dynamics would have to be Newtonian, which would contradictorily imply $g\approx\gN\lesssim\az$. This follows essentially from the basic tenets.\footnote{Strictly speaking, the second tenet says that dynamics is Newtonian everywhere if $g$ is much larger than $\az$ everywhere. Here I assume that this is so also in any sub region in the system.}
\par
However, in systems with $\gN\gg\az$, the maximum value that the excess MOND acceleration can attain depends on the fastness with which MOND tends to ND in the limit $\az\rar 0$. We can demonstrate this with the pristine, algebraic MOND relation between $g$ and $\gN$, which can be written as $g=\n(\gN/\az)\gN$, with $\n(y\ll 1)\approx y^{-1/2}$, $\n(y\gg1)\approx 1$.
For this case $(g-\gN)/\az=s(y)\equiv[\n(y)-1]y$, where $y=\gN/\az$. So, we are interested in the maximum value that $s(y)$ can attain. We see that if $\n-1$ vanishes slower than $\a/y$ (for some constant $\a$), $s$ has no upper bound. But, if $\n-1$ vanishes as $\a/y$ or faster, $s$ does have an upper bound.
This bound has to be of the order of 1 by our assumption of weak compatibility. As mentioned in Sec. \ref{additional}, constraints from the Solar system force on us `strong compatibility' of MOND with standard dynamics. This dictates that $|\vg-\vg\_N|/\az$ vanishes at high accelerations. This quantity must thus be bound by a number of the order of 1, which depends on the theory.
This prediction was first discussed in Brada \& Milgrom (1999a) in the framework of the nonlinear-Poisson formulation of MOND, and tested and confirmed for a sample of disc galaxies in Milgrom \& Sanders (2005).
\par
Let $\gN\^{max}$ be the maximal Newtonian acceleration in a system. One expects that if $\gN\^{max}\gtrsim\az$ then the maximal value of the excess MOND acceleration would be of the order of $\az$.
But, in a DML system, where $\gN\^{max}\ll\az$, the discussion in Sec. \ref{DML} implies that the maximum excess acceleration is expected to be
$g\^{max}\sim(\az\gN\^{max})^{1/2}$.
\par
Expressed as a property of a putative DM, this MOND prediction says that the acceleration produced by DM alone can never much exceed $\az$. There is
no known reason for such a constraint to hold in the DM paradigm.

\subsection{\label{maxac} $\S_M$ is a maximum and an accumulation value of the central surface densities of `dark haloes'}
Suppose we have a system of mass density $\r(\vr)$ whose Newtonian acceleration field is $\vg\_N(\vr)$. One then uses some techniques to map the dynamics of the system, namely, to measure some components of the accelerations of test particles at selected locations in the system, using, e.g., RCs, lensing, hot gas hydrostatics, etc.  These data are then fitted by an acceleration field, assumed to be derivable from a potential, by making further assumptions (for example that the excess acceleration is due to a spherical DM halo of a certain density profile). Call the MOND  prediction of this field $\vg(\vr)$ (mind again, MOND does not always predict the existence of an acceleration field; so here $\vg$ is the fit to the MOND prediction). If one attributes the excess
$\vg_h=\vg-\vg\_N$ to DM, then the density of this `PM' is $\r\_p(\vr)=-(\fpg)^{-1}\div\vg_h$.
\par
Let $g\_{max}$ be the maximal characteristic value of $|\vg_h|$ in the system, and $r_0$ the characteristic size of the `halo': the radius at which $|\vg_h|$ drops appreciably from $g\_{max}$. Then the characteristic phantom density is $\langle\rp\rangle\sim -(\fpg)^{-1}\langle\div\vg_h\rangle\sim (\fpg)^{-1}g\_{max}/r_0$. So, the characteristic mean surface density of the `halo' is $\langle\S_p\rangle\sim 2\langle\rp\rangle r_0\sim g\_{max}/2\pi G$. Seen somewhat differently, if we assume a quasi-spherical geometry, then $\rp\approx -(\fpg)^{-1}r^{-2}d(r^2g^r_h)/dr=-(\fpg)^{-1}(dg^r_h/dr+2g^r_h/r)$, where $g^r_h$ is the radial component of $\vg_h$, ($|g^r_h|=|\vg_h|$). The two terms are opposite in sign; so if we approximate $g^r_h/r\approx -dg^r_h/dr$, and integrate we get the same result for $\langle\S_p\rangle$.
This relation
\beq \langle\S_p\rangle\sim  \frac{g\_{max}}{\az}\sm\equiv \xi\S_M, \eeqno{sisisi}
 is a result of the definitions alone.
\par
Now, according to section \ref{maximum}, for all systems with $\gN\gtrsim\az$, which is an accumulation value for astrophysical systems, we have $g\_{max}\sim\az$, in which case $\xi\sim 1$.
\par
For DML systems we saw that $g\_{max}\sim(\az\gN^{max})^{1/2}$; so,
\beq \langle\S_p\rangle=\bar\xi(\gN^{max}/\az)^{1/2}\sm, \eeqno{mioapl}
with $\bar\xi\sim 1$. The coefficients $\xi,~\bar\xi\sim 1$ depend somewhat on the particular theory (including details, such as the form of the relevant interpolating function), and on the exact mass distribution $\r$.
However, this leeway in the values of $\xi$ and $\bar\xi$ is small compared with the relative ranges over which system parameters (masses, sizes, etc.) vary.
\par
This prediction, which concerns a pure `halo' property,
was discussed in detail in Milgrom (2009b), in the context of the MG formulation based on the nonlinear Poisson equation, including explicit determination of $\xi$ and $\bar\xi$.\footnote{From the discussion in Appendix B.2 in Milgrom (2009c), we see that the same results hold in QUMOND.} This analysis was instigated by the finding of Donato \& al. (2009), of  a `universal' halo surface density (which turns out to be very nearly $\sm$). The ubiquity of this value was already evident in the analysis of Milgrom \& Sanders (2005) that considered a sample of mostly non DML disc galaxies.
Additional relevant data and analysis can be found in Salucci \& al. (2012).
\par
A related finding pertaining to the mean {\it baryonic} surface density within one `halo' scale length -- which thus constitute a relation between the visible and `dark' matter -- was discussed in Gentile \& al. (2009).
\par
If interpreted in the framework of DM, this MOND prediction pertains to pure halo properties. Unlike prediction (\ref{asym}), which concerns the asymptotics of the `halo', the present prediction concerns its bulk properties. Such a critical halo surface density is not reproduced in the DM paradigm.

\subsection{\label{rcs}Rotation curves of disc galaxies}
The prediction of RCs of disc galaxies does not count among the MOND laws. Unlike these laws, which each pertains to a very limited, one- or two-parameter aspect of galactic dynamics, the RC measures the full midplane dynamics (at least for circular orbits). The RC prediction is rather more elaborate, and, in fact, subsumes several of the MOND laws: (\ref{asym}), (\ref{btfr}), and (\ref{bound}). Still, in the present context it is interesting to note that even for RC prediction, the room left by the basic tenets  for differences between theories is only in the details, not in the essentials.
\par
This can be seen as follows:
two radii characterize the dynamics of a galaxy in MOND: $R$, the radius containing most of the mass, and $\RM$.
We saw that the basic tenets fully dictate the RC -- in shape and magnitude -- far beyond the larger of the two (the asymptotic regime). Now, if $R<\RM$, most of the matter is in the Newtonian regime. The basic tenets then dictate a Newtonian -- hence theory independent -- RC within $R$, or even far beyond it, if $R\ll\RM$ (except, perhaps, for a small region near the very centre).\footnote{Depending on the density distribution near the centre, the accelerations may drop below $\az$ there.} So theories can differ only in the interpolating region just below and just above $\RM$.
\par
When $\RM<R$ (strictly speaking $\RM\ll R$), the galaxy is fully in the MOND regime. In this case, characterize the disc galaxy, most generally, by a mass $M$, $R$, and dimensionless structure parameters $\eta_i$ (shape parameters, mass ratios of different components, etc.). Then, dimensional considerations and SI dictate that the MOND rotational speed in the midplane, $V(r)$, can be written, for any MOND theory, in the form

  \beq V(r)= [\az r V^2\_N(r)]^{1/4}q(r/R,\eta_i), \eeqno{lalamaka}
where $V\_N(r)$ is the Newtonian rotational speed.\footnote{Note that while $V\_N$, which is defined as $V\_N=(rg\_N)^{1/2}$, has dimensions of velocity it does not transform like a velocity under scaling, since $g\_N$ scales like $\l^{-2}$. It is the quantity $rV^2\_N$ that scales appropriately (i.e. it is SI).}
The dimensionless $MG/R^2\az=(\RM/R)^2$ cannot appear as an argument of the dimensionless and SI $q$ because it is not SI.
Furthermore, $q\approx 1$ for $r\gg R$, as dictated by the basic tenets, since the prefactor tends to $\vinf=(M\azg)^{1/4}$ there.
This is as far as we can take the basic tenets on the issue of RCs for DML galaxies.
\par
In MI theories, for circular orbits in the midplane of a disc, $g=V^2/r$ is necessarily a function of the local value $\gN$ (Milgrom 1994). Then, for DML discs we have $g=(\az\gN)^{1/2}q^2$; this implies that $q\equiv 1$, irrespective of the mass distribution, since the only function of $\gN$ with the required properties is a constant (e.g., $\gN/\az$ is not SI).
\par
Some numerical comparisons of DML RCs between MI and existing MG  theories for several disc models were shown in Brada \& Milgrom (1995). More generally, Milgrom (2012b) showed that the rms velocities predicted by these two classes in the DML differ by no more than 5 percent. The difference is, however, systematic [$V(MI)>V(MG)$] and could be used to discriminate observationally between the theory classes.

\section{Discussion}
I have listed and discussed a number of predicted MOND laws that should underlie galaxy dynamics. The list comprises exact relations, correlations, and constraints that the attributes of galaxies and galactic systems should satisfy, according to MOND. These laws are independent in the sense that they do not follow from each other in the DM paradigm. Some of them involve $\az$ in various roles, in laws pertaining to baryons alone, to `DM' alone, or to relations between the two components. In all deep-MOND results, $\az$ appears in the combined constant $\azg=G\az$. In other laws, $\az$ appears as a critical surface density $\sm=\az/2\pi G$, but still in different roles, e.g., as a property of baryons in law (\ref{siglimit}), and as a property of `DM' in law (\ref{centralsig}).
\par
I showed that to a large extent these laws follow from only the basic tenets of MOND. But there are exceptions, where some of the laws, or partial aspects of them, require additional assumptions. Such assumptions are, by and large, satisfied by existing MOND theories, but it is not necessary that any MOND theory should satisfy them.
\par
For example, laws (\ref{siglimit}) and (\ref{msigma}) do not apply to high-acceleration systems, which, in turn, cannot be quasi isothermal, such as stars, or compact stellar systems. And, we found that the exact validity and form of laws (\ref{efe}) and (\ref{disc}) are theory dependent.
\par
There are, of course, many aspects of galaxy dynamics that are not dictated by the basic tenets, and require a detailed theory to predict. For example the details of the gravitational field of a galaxy, and even details of its RC near the Newtonian-MOND transition, depend on the exact theory (for examples, see Zhao \& Famaey 2010, Milgrom 2012b, L\"{u}ghausen \& al. 2013).

\clearpage


\begin{thebibliography}{}
\bibitem[Babichev \& al. (2011)]{babichev11}Babichev, E.,  Deffayet, C., \&  Esposito-Farese, G. 2011, Phys. Rev. D 84, 061502
\bibitem[Bekenstein (2004)]{bek04}Bekenstein, J.D. 2004, Phys. Rev. D 70, 083509
\bibitem[Bekenstein (2011)]{bek11}Bekenstein, J. 2011, S´eminaires de l'IAP,
[www.iap.fr/activites/seminaires/IAP/2011/Presentations/bekenstein.pdf]
\bibitem[Bekenstein \& Milgrom (1984)]{bm84}Bekenstein, J. \&  Milgrom, M. 1984, ApJ. 286, 7
\bibitem[Brada \& Milgrom (1995)]{brada95}Brada, R. \& Milgrom, M. 1995, MNRAS 276, 453
\bibitem[Brada \& Milgrom (1999a)]{brada99a}Brada, R. \& Milgrom, M. 1999a, ApJ 512, L17
\bibitem[Brada \& Milgrom (1999b)]{brada99b}Brada, R. \& Milgrom, M. 1999b, ApJ 519, 590
\bibitem[Christodoulou (1991)]{christ91}Christodoulou, D.M. 1991, ApJ 372, 471
\bibitem[Donato \& al. (2009)]{donato09}Donato, F., et al., 2009, MNRAS 397, 1169
\bibitem[Famaey \& McGaugh (2012)]{fm12}Famaey, B \& \& McGaugh, S. 2012,  Living Rev.  Relativ. 15, 10
\bibitem[Gentile \& al. (2009)]{gentile09}Gentile, G., Famaey, B., Zhao, H.S, \& Salucci, P. 2009, Nature 461, 627
\bibitem[L\"{u}ghausen  \& al.(2013)]{lueghausen13}L\"{u}ghausen, F., Famaey, B., Kroupa, P., Angus, G., Combes, F., Gentile, G., Tiret, O., \& Zhao, H.  2013, MNRAS 432, 2846
\bibitem[McGaugh (2011)]{mcgaugh11a}McGaugh, S.S. 2011, Phys. Rev. Lett. 106, 121303
\bibitem[McGaugh (2012)]{mcgaugh12}McGaugh, S.S. 2012, AJ 143, 40
\bibitem[McGaugh \& Milgrom (2013a)]{mm13a}McGaugh, S. \&  Milgrom, M. 2013a, ApJ 766, 22
\bibitem[McGaugh \& Milgrom (2013b)]{mm13b}McGaugh, S. \&  Milgrom, M. 2013b, ApJ 775, 139
\bibitem[Milgrom (1983)]{milgrom83}Milgrom, M. 1983, ApJ 270, 365
\bibitem[Milgrom (1984)]{milgrom84}Milgrom, M. 1984, ApJ 287, 571
\bibitem[Milgrom (1986)]{milgrom86}Milgrom, M. 1986, ApJ 306, 9
\bibitem[Milgrom (1989a)]{milgrom89a}Milgrom, M. 1989a, ApJ 338, 121
\bibitem[Milgrom (1989b)]{milgrom89b}Milgrom, M. 1989b, AA 211, 37
\bibitem[Milgrom (1994)]{milgrom94}Milgrom, M. 1994, Ann. Phys. 229, 384
\bibitem[Milgrom (1997)]{milgrom97}Milgrom, M. 1997, Phys. Rev. E 56, 1148
\bibitem[Milgrom (2001)]{milgrom01}Milgrom, M. 2001, MNRAS 326, 1261
\bibitem[Milgrom (2002)]{milgrom02}Milgrom, M. 2002, ApJL 577, L75
\bibitem[Milgrom (2008)]{milgrom08}Milgrom, M. 2008,  XIX Rencontres de Blois, arXiv:0801.3133
\bibitem[Milgrom (2009a)]{milgrom09a}Milgrom, M. 2009a, ApJ 698, 1630
\bibitem[Milgrom (2009b)]{milgrom09b}Milgrom, M. 2009b, MNRAS 398, 1023
\bibitem[Milgrom (2009c)]{milgrom09c}Milgrom, M. 2009c, MNRAS 399, 474
\bibitem[Milgrom (2010a)]{milgrom10a}Milgrom, M. 2010a, MNRAS 403, 886
\bibitem[Milgrom (2010b)]{milgrom10b}Milgrom, M. 2010b, MNRAS 405, 1129
\bibitem[Milgrom (2011)]{milgrom11}Milgrom, M. 2011, Acta Physica Polonica B vol. 42, 2175
\bibitem[Milgrom (2012a)]{milgrom12a}Milgrom, M. 2012a, Phys. Rev. Lett. 109, 131101
\bibitem[Milgrom (2012b)]{milgrom12b}Milgrom, M. 2012b, Phys. Rev. Lett. 109, 251103
\bibitem[Milgrom (2013a)]{milgrom13a}Milgrom, M. 2013a, Phys. Rev. Lett. 111, 041105
\bibitem[Milgrom (2013b)]{milgrom13b}Milgrom, M. 2013b, arXiv:1311.2579
\bibitem[Milgrom \& Sanders (2005)]{ms05}Milgrom, M. \& Sanders, R.H. 2005,  MNRAS 357, 45
\bibitem[Milgrom \& Sanders (2008)]{ms08}Milgrom, M. \& Sanders, R.H. 2008, ApJ 678, 131
\bibitem[Ostrogradski (1850)]{ostro1850}Ostrogradski, M. 1850, Mem. Ac. St. Petersbourg VI 4, 385
\bibitem[Salucci \& al. (2012)]{salucci12}Salucci, P., Wilkinson, M.I., Walker, M.G., Gilmore, G.F.,  Grebel, E.K., Koch, A., Frigerio Martins, C., \& Wyse, R.F.G. 2012, MNRAS 420, 2034
\bibitem[Sanders (2010)]{sanders10}Sanders, R.H. 2010, MNRAS 407, 1128
\bibitem[Scarpa (2006)]{scarpa06}Scarpa, R. 2006, AIP Conf. Proc. 822, 253, arXiv:astro-ph/0601478
\bibitem[Tiret \& Combes (2007a)]{tiret07a}Tiret, O. \& Combes, F. 2007a, AA 464, 517
\bibitem[Tiret \& Combes (2007b)]{tiret07b}Tiret, O. \& Combes, F. 2007b  arXiv:0709.3376
\bibitem[Woodard (2007)]{woodard07}Woodard, R.P. 2007, Lect. Notes Phys. 720, 403
\bibitem[Wu \& al. (2008)]{wu08}Wu, X., Famaey, B., Gentile, G., Perets, H. \& Zhao, H.S. 2008, MNRAS 386, 2199
\bibitem[Zhao \& Famaey (2010)]{zf10}Zhao, H.S. \&  Famaey, B. 2010, Phys. Rev. D 81, 087304
\bibitem[Zhao \& Famaey (2012)]{famaey12}Zhao, H.S. \&  Famaey, B. 2012, Phys. Rev. D86, 067301

\end{thebibliography}
\end{document}